\documentclass[preprint,aps]{revtex4}
\usepackage[dvips]{graphicx}

\begin{document}

\textheight 9.5in

\title{ Test of the Littlest Higgs model through the correlation \\
among $W$ boson, top quark and Higgs masses}

\author{Sin Kyu Kang}\email{skkang@snut.ac.kr} \affiliation{School of Liberal Arts, Seoul
 National University of Technology, Seoul 139-743, Korea}
\author{C.~S.~Kim}\email{cskim@yonsei.ac.kr} \affiliation{Department
  of Physics and IPAP, Yonsei University, Seoul 120-749, Korea}
\author{Jubin Park}\email{jbpark@cskim.yonsei.ac.kr} \affiliation{Department
  of Physics and IPAP, Yonsei University, Seoul 120-749, Korea}

%
%
\begin{abstract}

\noindent Motivated by the recent precision measurements of the $W$
boson mass and top quark mass, we test the Littlest Higgs model by
confronting the prediction of $M_{_W}$  with the current and
prospective measurements of $M_{_W}$ and $M_{t}$ as well as through
the correlation among $M_{_W}$, $M_{t}$ and Higgs mass.
We argue that the
current values and accuracy of $M_{_W}$ and $M_{t}$ measurements
tend to favor the Littlest Higgs model over the standard
model, although the most recent electroweak data may appear to be
consistent with the standard model prediction. In this analysis,
the upper bound on the global $SU(5)$ symmetry breaking scale
turned out to be 26.3 TeV. We also discuss how the masses of the
heavy gauge boson $M_{B^{'}}$ in the Littlest Higgs model can be
predicted from the constraints on the model parameters.
\\

 \end{abstract}

\maketitle

\section{Introduction}
\label{sec:1}

There have been a great deal of works on the precision test of the
standard model (SM) because of the incredibly precise data obtained
at the LEP and the new measurements of $M_{_W}$ and $M_{t}$ at the
Fermilab Tevatron~\cite{:2007ypa,Group:2008nq} as well as the recent
theoretical progress in the higher order radiative
corrections~\cite{Buttar:2006zd}.
With such a dedicated effort for a long time to test the SM, it has
been confirmed that the SM is the right model to describe the
electroweak phenomena at the current experimental energy scale. What
remains elusive is the origin of the electroweak symmetry breaking
for which a Higgs boson is responsible in the SM. It has been known
for some time that radiative corrections in the SM exhibit a small
but important dependence on the Higgs boson mass, $M_{h}$. As a
result, the value of $M_{h}$ can, in principle, be predicted by
comparing a variety of precision electroweak measurements with one
another. The recent global fits to all precision electroweak data
(see J. Erler and P. Langacker~\cite{PDG}) lead to $M_{h} =
113^{+56}_{-40}$ ($1\sigma$ confidence level (CL)) and $M_{h} < 241$ GeV ($95\%$ CL).
Those constraints are very consistent with bounds from direct
searches for the Higgs boson at LEPII via $e^{+}+e^{-}
\longrightarrow Z h$, $M_{h} > 114.4$ GeV~\cite{Barate:2003sz}.
Together, they seem to suggest the range, 114 GeV $< M_{h} <$ 241
GeV, and imply very good consistency between the SM and experiment.
However, in the context of the SM valid all the way up to the Planck
scale, $M_{h}$ diverges due to a quadratic divergence at one loop
level unless it is unnaturally fine-tuned. Thus, we need a new
physics beyond the SM to stabilize $M_{h}$, which
is a so-called  hierarchy problem that has motivated the construction
of the LHC. Candidates for this physics include supersymmetry and
technicolor models relying on strong dynamics to achieve electroweak
symmetry breaking.

Inspired by {\it dimensional
deconstruction}~\cite{ArkaniHamed:2001ca}, an intriguing alternative
possibility that the Higgs boson is a pseudo Goldstone boson
~\cite{Georgi:1974yw,ArkaniHamed:2002qy}
has been revived by Arkani-Hamed {\it et al.}. They showed that the
gauge and Yukawa interactions of the Higgs boson can be incorporated
in such a way that a quadratically divergent one-loop contribution
to $M_{h}$ is canceled. The cancelation of this contribution occurs
as a consequence of the special ¡°collective¡± pattern in which the
gauge and Yukawa couplings break the global symmetries. Since the
remaining radiative corrections to $M_{h}$ are much smaller, no fine
tuning is required to keep the Higgs boson sufficiently light if the
strong coupling scale is of order 10 TeV. Such a light Higgs boson
was called ``little Higgs". The models with little Higgs are
described by nonlinear sigma models and trigger electroweak symmetry
breaking by the collective symmetry breaking mechanism. Many such
models with different ``theory space" have been constructed
~\cite{ArkaniHamed:2002qy,Schmaltz:2004de},
and  electroweak  precision constraints on various little Higgs
models have been investigated by performing global fits to the
precision data
~\cite{Csaki:2002qg,Csaki:2003si,Deandrea:2004qq}.
It is worthwhile to notice that the little Higgs models generally
have three significant scales: an electroweak scale $v \sim
\frac{g^{2}f}{4\pi} \sim$  200 GeV, a new physics scale $g \cdot f$
$\sim$ 1 TeV and a cut-off scale of the non-linear sigma model
$\Lambda$ $\sim$ $4\pi f$ $\sim$ 10 TeV, where $f$ is the scale of
the global symmetry breaking. Therefore, we expect that the little
Higgs  models have rich and distinguishable TeV scale phenomena
unlike other models, which provides strong motivation to probe them
at the LHC.

Very recently, Fermilab CDF collaboration has reported the most precise single measurement
of the $W$ boson mass to date from Run II of the Tevatron ~\cite{:2007ypa},
\begin{eqnarray}
M_{_W}=80.413\pm 0.048 ~\mbox{GeV},
\end{eqnarray}
and updated the world average \cite{Alcaraz:2007ri} to
\begin{eqnarray}
M_{_W}=80.398\pm 0.025 ~\mbox{GeV}.
\end{eqnarray}
In addition, the world average result of $M_{t}$ from the Tevatron
experiments CDF and D0 has been given~\cite{Group:2008nq} by
\begin{eqnarray}
M_{t}=172.6 \pm 1.4~\mbox{GeV}.
\end{eqnarray}

The mass of the top quark is now known with a relative precision of
$0.8\%$, limited by the systematic uncertainties, and can be
reasonably expected that with the full Run-II data set the top-quark
mass will be known to much better than $0.8\%$ in the foreseeable
future. With the current level of experimental uncertainties as well
as prospective sensitivities on $M_{_W}$ and $M_{t}$, we are
approaching to the level to test the validity of new physics beyond
the SM by a direct comparison with data or to strongly constrain new
physics models.
%
%
%

The correlation among $M_{t}$, $M_{_W}$ and $M_{h}$ is an important
prediction of the SM, and thus deviations from it should be
accounted for by the effects of new physics. In the minimal
supersymmetric standard model (MSSM) case, the allowed ranges for
$M_{_W}$ and $M_{t}$ were checked by considering various parameter
spaces of the MSSM~\cite{Heinemeyer:2006nm}. They showed that the
previous experimental results for $M_{_W}$ and $M_{t}$ tend to favor
the MSSM over the SM. Motivated by this fact, in this letter,
we confront the Littlest Higgs model (LHM)~\cite{ArkaniHamed:2002qy}
with more precision measurements of $M_{_W}$ and $M_{t}$ than
before by computing the prediction of $M_{_W}$ in the LHM. We
examine whether the current precision measurements of $M_{_W}$ and
$M_{t}$ tend to favor the LHM over the SM or not. From the careful
numerical analysis, we obtain some constraints on the model
parameters such as the global $SU(5)$ symmetry breaking scale and
the mixing angles between heavy gauge bosons. By using the
constraints on the model parameters, we show how the mass of heavy
gauge boson $B^{\prime}_{\mu}$ can be predicted, which could be
probed at the LHC.

The organization of this letter is as follows. In Sec. II we briefly
review the LHM. In Sec. III we discuss how the formula for $M_{_W}$
can be derived from the effective
 theory of the LHM, and confront the prediction of $M_{_W}$
 with the current and prospective measurements of $M_{_W}$ and $M_{t}$.
 We also show how
  an upper bound on the global symmetry breaking scale $f$ can be obtained and how it is correlated
  with the Higgs mass.
  In Sec. IV  we investigate how the mixing parameters in the LHM
  can be constrained, and discuss how the mass of the heavy gauge
  boson $B^{'}_{\mu}$ in the LHM
  can be predicted from the constraints on the model parameters.
Finally we conclude our work.

\section{Aspects of the littlest Higgs model}
\label{sec:2}

We start with reviewing the aspects of the LHM which are relevant to
our work. The LHM is one of the simplest and phenomenologically
viable models, which realizes little Higgs idea.
It initially has a global symmetry $SU(5)$ which is broken down
to a global symmetry $SO(5)$ via a vacuum expectation value of order $f$, and
 a gauge group $[SU(2)\times U(1)]^{2}$ which is broken down to $SU(2)\times U(1)$, identified as
 the electroweak gauge symmetry. The characteristic feature of the LHM is to predict the
 existence of the new gauge bosons with masses of order TeV.
The vacuum expectation value (VEV) associated with the spontaneous global symmetry breaking of $SU(5)$ is proportional
to the $5\times 5$ symmetric matrix $\Sigma_0$ given by
\begin{equation}
\Sigma_{0}=\left(
  \begin{array}{ccccc}
     &  &  & 1 &  \\
     &  &  &  & 1 \\
     &  & 1 &  &  \\
    1 &  &  &  &  \\
     & 1 &  &  &  \\
  \end{array}
\right).
\end{equation}

The global symmetry breaking yields 14 Goldstone bosons which
transform under the electroweak $SU(2)$ symmetry as a real singlet,
a real triplet, a complex doublet and a complex triplet:
\begin{equation}
\mathbf{14}=\mathbf{1}_{0}+\mathbf{3_{0}}+\mathbf{2_{\pm_{1/2}}}+\mathbf{3_{\pm_{1/2}}}~.
\end{equation}
Among them four massless Goldstone bosons, $\mathbf{1}_{0}$ and
$\mathbf{3_{0}}$ are eaten by the gauge fields so that the gauge
symmetry $[SU(2)\times U(1)]^{2}$ is broken down to its diagonal
subgroup $SU(2)\times U(1)$. The remaining complex doublet
$\textbf{2}_{\pm 1/2}$ and triplet $\textbf{3}_{\pm 1/2}$ are
identified as a component of the SM Higgs sector and an extra
complex triplet Higgs, respectively. The generators of the gauge
symmetry embedded into $SU(5)$ are given by
 \begin{equation}
 Q^{a}_{1}=\left(
             \begin{array}{ccc}
               \sigma^{a}/2 & 0 & 0 \\
               0 & 0 & 0 \\
               0 & 0 & 0 \\
              \end{array}
           \right),~~~
  Y_{1}={\rm diag}(-3,-3,2,2,2)/10,
 \end{equation}
 \begin{equation}
 Q^{a}_{2}=\left(
             \begin{array}{ccc}
               0 & 0 & 0 \\
               0 & 0 & 0 \\
               0 & 0 & -\sigma^{a*}/2 \\
             \end{array}
           \right),~~~~
  Y_{2}={\rm diag}(-2,-2,-2,3,3)/10,
 \end{equation}
 where $\sigma^{a}$ are the Pauli spin matrices and $Q^{a}_i$ and $Y_{i}$ are each $SU(2)$ and $U(1)$ generators, respectively.
 Then, the generators of the electroweak symmetry $SU(2)_L\times U(1)_Y$ are
 $Q^a=(Q^a_1+Q^a_2)/\sqrt{2}$ and $Y=Y_1+Y_2$.

The fluctuations of the remaining Goldstone bosons in the broken direction can be described by
$\Pi=\pi^{a}X^{a}$ with the broken generators of the $SU(5)$, $X^{a}$ .
Then the Goldstone bosons can be parameterized by a nonlinear sigma model field $\Sigma(x)$,
\begin{equation}
\Sigma(x)=e^{i\Pi/f}\Sigma_{0}e^{i\Pi^{T}/f}=e^{2i\Pi/f}\Sigma_{0}.
\end{equation}
In terms of uneaten fields, the Goldstone boson field, $\Pi$, is
given by
\begin{equation}
 \Pi=\left(
             \begin{array}{ccc}
               0 & \frac{H^{\dagger}}{\sqrt{2}} & \Phi^{\dagger} \\
                \frac{H}{\sqrt{2}}& 0 & \frac{H^{\ast}}{\sqrt{2}} \\
               \Phi & \frac{H^{T}}{\sqrt{2}} & 0
             \end{array}
           \right),
 \end{equation}
where $H$ denotes the little Higgs doublet $(h^0, h^{\dagger})$ and $\Phi$ is a complex
triplet scalar field. We note that the triplet scalar field $\Phi$ should have a small
expectation value of order GeV in order to not give too large contribution to the $T$ parameter
\cite{Csaki:2002qg}.

The kinetic energy term of the nonlinear sigma field $\Sigma$ is given by
\begin{equation}
\frac{f^{2}}{8}Tr D_{\mu}\Sigma \cdot (D^{\mu}\Sigma)^{\dagger},
\end{equation}
where the covariant derivative of $\Sigma$ is
\begin{equation}
D_{\mu}\Sigma=\partial_{\mu}\Sigma -
i\Sigma_{j}[g_{j}W^{a}_{j\mu}(Q^{a}_{j}\Sigma + \Sigma
Q^{a\,T}_{j})+g^{'}_{j}B_{j\mu}(Y_{j}\Sigma+\Sigma Y_{j})], \label{kinetic}
\end{equation}
with $j=1,2$. Here $W^{a}_{j\mu}$ and $B_{j\mu}$ stand for the $SU(2)$ and
$U(1)$ gauge fields, respectively and $g_j$ and $g^{\prime}_j$ denote
the corresponding gauge coupling constants.

It is convenient to expand $\Sigma$ around the VEV in powers of $1/f$,
\begin{eqnarray}
\Sigma=\Sigma_{0}&+&\frac{2i}{f}\left(
             \begin{array}{ccc}
             \Phi^{\dagger} & \frac{H^{\dagger}}{\sqrt{2}} & \mathrm{0}_{2\times 2} \\
             \frac{H^{*}}{\sqrt{2}} & 0 & \frac{H}{\sqrt{2}} \\
             \mathrm{0}_{2\times2} & \frac{H^{T}}{\sqrt{2}} & \Phi
             \end{array}
\right) \nonumber \\
&-&\frac{1}{f^{2}}\left(
             \begin{array}{ccc}
             H^{\dagger}H^{*} & \sqrt{2}\Phi^{\dagger}H^{T} & H^{\dagger}H+2\Phi^{\dagger}\Phi\\
             \sqrt{2}H\Phi^{\dagger} & 2HH^{\dagger} & \sqrt{2}H^{*}\Phi \\
             H^{T}H^{*}+2\Phi\Phi^{\dagger} & \sqrt{2}\Phi H^{\dagger} & H^{T}H
             \end{array}
\right)  \label{expand} \\
&+&\textit{O}\left(\frac{1}{f^{3}}\right). \nonumber
\end{eqnarray}
Inserting Eq. (\ref{expand}) into Eq. (\ref{kinetic}), we obtain the mixing terms between gauge bosons as follows,
\begin{eqnarray}
\mathcal{L}_{\Sigma,~\rm LO} & \sim & \frac{f^{2}}{8} \, \mathrm{Tr} | \,
\Sigma_{j=1,2} [ \,
g_{j}W_{j\,\mu}^a(Q_{j}^a\Sigma_{0}+\Sigma_{0}Q^{aT}_{j})
+g^{'}_{j}B_{j\,\mu}(Y_{j}\Sigma_{0}+\Sigma_{0}Y_{j})] \, |^{2} \nonumber \\
&\sim &  \frac {f^{2}}{8} \{ \left(g_{1}^{2}W_{1\,\mu}^aW_{2}^{a\mu}-2g_{1}g_{2}W_{1\,\mu}^aW_{2}^{a\mu}+g_{2}^{2}W_{2\,\mu}^aW_{2}^{a\mu}\right)\\
&&+\frac{1}{5}\left(\,g_{1}^{'2}B_{1\,\mu}B_{1}^{\mu}-2g_{1}^{'}g_{2}^{'}B_{1\,\mu}B_{2}^{\mu}+g_{2}^{'2}B_{2\,\mu}B_{2}^{\mu}\right)\}. \nonumber
\end{eqnarray}
With the help of the following transformations
\begin{equation}
W^{a}_{\mu}=sW^{a}_{1\,\mu}+cW^{a}_{2\,\mu}~,~~~ W^{a'}_{\mu}=-cW^{a}_{1\,\mu}+sW^{a}_{2\,\mu},
\end{equation}
\begin{equation}
B_{\mu}=s^{'}B_{1\,\mu}+c^{'}B_{2\,\mu}~,~~~
B^{'}_{\mu}=-c^{'}B_{1\,\mu}+s^{'}B_{2\,\mu},
\end{equation}
with
\begin{equation}
s=\frac{g_{2}}{\sqrt{g_{1}^{2}+g_{2}^{2}}}~,~~~
c=\frac{g_{1}}{\sqrt{g_{1}^{2}+g_{2}^{2}}},
\end{equation}
\begin{equation}
s^{'}=\frac{g_{2}^{'}}{\sqrt{g_{1}^{'2}+g_{2}^{'2}}}~,~~~
c^{'}=\frac{g_{1}^{'}}{\sqrt{g_{1}^{'2}+g_{2}^{'2}}},
\end{equation}
two massive states $W^{a\prime}_{\mu}$
and $B^{\prime}_{\mu}$ are obtained whose masses are given by
\begin{eqnarray}
M_{W^{a'}_{\mu}}=\sqrt{g_{1}^{2}+g_{2}^{2}}
\, \frac{f}{2}, \nonumber \\
M_{B^{'}_{\mu}}=\sqrt{g_{1}^{'2}+g_{2}^{'2}}\frac{f}{\sqrt{20}},
\end{eqnarray}
respectively, and two massless $W^a_{\mu}$ and $B_{\mu}$ bosons which are
identified as the massless SM gauge bosons before the electroweak
symmetry breaking. Those SM gauge fields become massive after the
electroweak symmetry breaking at a few hundred GeV scale.
Hereafter we denote the SM gauge fields in the mass basis
as $W,Z$ and $A$.
We also notice that the SM gauge couplings are $g=g_1s=g_2c$ and
$g^{\prime}=g^{\prime}_{1}\,s^{\prime}=g^{\prime}_{2}\,c^{\prime}$
for $SU(2)_{L}$ and $U(1)_{Y}$, respectively.

\section{Prediction of $M_{_W}$ and upper bound on $f$ }

The primary goal of our work is to estimate the prediction for the
mass of $W$ boson in the LHM. To do this, it is convenient to
construct low energy effective lagrangian for the LHM below the mass
scales of the heavy gauge bosons and then extract the corrections
coming from higher dimensional operators. The quartic couplings of
the Higgs and gauge bosons can be obtained by expanding the
next-to-leading order terms of the non-linear sigma field in the
kinetic term,
\begin{eqnarray}
\mathcal{L}_{\Sigma,~\rm NLO} &\sim& \frac{1}{2} \mathrm{Tr} | \,
\Sigma_{j=1,2}
[\,g_{j}W_{j\mu}^a(Q_{j}^a\Pi\Sigma_{0}+\Pi\Sigma_{0}Q^{aT}_{j})
+g^{'}_{j}B_{j\mu}(Y_{j}\Pi\Sigma_{0}+\Pi\Sigma_{0}Y_{j})] \, |^{\,2}.
\end{eqnarray}
Expressing these gauge bosons in terms of the mass eigenstates $W_{\mu}^a,~ W^{a'}_{\mu},~ B_{\mu}$
and $B^{'}_{\mu}$, the quartic terms are given by
\begin{eqnarray}
\mathcal{L}_{\Sigma,~\rm NLO} &\sim& +\frac{1}{4}g^{2}\left(W_{\mu}^{a}W^{b\,\mu}-
\frac{(c^{2}-s^{2})}{sc}W_{\mu}^{a}W^{\prime\,b\,\mu}\right)
\mathrm{Tr}[H^{\dagger}H\delta^{ab}+2\Phi^{\dagger}\Phi\,\delta^{ab}+2\sigma^{a}\Phi^{\dagger}\sigma^{b\,T}\Phi]\nonumber \\
&&-\frac{1}{4}g^{2}\left(W_{\mu}^{\prime\,a}W^{\prime\,a\,\mu}\,\mathrm{Tr}[H^{\dagger}H+
2\Phi^{\dagger}\Phi]
-\frac{(c^{4}+s^{4})}{2s^{2}c^{2}}W_{\mu}^{\prime\,a}W^{\prime\,b\mu}\,\mathrm{Tr}[2\sigma^{a}\Phi^{\dagger}
\sigma^{bT}\Phi]\right)\nonumber \\
&&+g^{\prime\,2}\left(B_{\mu}B^{\mu}-\frac{(c^{\prime\,2}-s^{\prime\,2})}{s^{\prime}c^{\prime}}B_{\mu}B^{\prime\,\mu}\right)
\,\mathrm{Tr}[\frac{1}{4}H^{\dagger}H+\Phi^{\dagger}\Phi]\nonumber \\
&&-g^{\prime\,2}\left(B^{\prime}_{\mu}B^{\prime\,\mu}\,\mathrm{Tr}[\frac{1}{4}H^{\dagger}H]
-\frac{(c^{\prime\,2}-s^{\prime\,2})^{2}}{4s^{\prime\,2}c^{\prime\,2}}B^{\prime}_{\mu}B^{\prime\,\mu}
\mathrm{Tr}[\Phi^\dagger\Phi]\right)+\dots~.
\end{eqnarray}
Integrating out the heavy gauge bosons $W^{a'}_{\mu}$ and $B^{'}_{\mu}$, we obtain additional operators
which cause modification of
relations between the SM parameters, and thus their coefficients can be constrained from electroweak precision data.
Among the additional operators, the terms quadratic with respect to the light gauge fields are given in the unitary gauge by
\begin{eqnarray}
\mathcal{L}_{~\rm effective} &\sim&-\frac{g^{2}(s^{2}-c^{2})^{2}}{8f^{2}}W^{a\,\mu}W^{a}_{\mu}h^{4}
-\frac{5g^{2}(s^{\prime\,2}-c^{\prime,2})^2}{8f^{2}}W^{3\mu}W^{3}_{\mu}h^{4} \nonumber \\
&&-\frac{g^{\prime\,2}(s^{2}-c^{2})^{2}}{8f^{2}}B^{\mu}B_{\mu}h^{4}
-\frac{5g^{\prime\,2}(s^{\prime\,2}-c^{\prime\,2})^{2}}{8f^{2}}B^{\mu}B_{\mu}h^{4}\nonumber \\
&&+\frac{gg^{\prime}(s^{2}-c^{2})^{2}}{4f^{2}}W^{3\mu}B_{\mu}h^{4} \nonumber \\
&&+\frac{g^{2}}{4f^{2}}W^{a\mu}W^{a}_{\mu}h^{4}+\frac{g^{\prime\,2}}{4f^{2}}B^{\mu}B_{\mu}h^{4}-\frac{gg^{\prime}}{2f^{2}}B^{\mu}W^{3}_{\mu}h^{4}
\nonumber \\
&&+\frac{g^{2}}{2}W^{a\mu}W^{a}_{\mu}\varphi^{2}+\frac{g^{2}}{2}W^{3\mu}W^{3}_{\mu}\varphi^{2}
+g^{\prime\,2}B^{\mu}B_{\mu}\varphi^{2}-2gg^{\prime}B^{\mu}W^{3}_{\mu}\varphi^{2}, \label{eff}
\end{eqnarray}
where we only take the  $h~\equiv{\rm Re}~h^{0}$ component of Higgs field $H$
and $\varphi~\equiv{\rm Re}~\phi^{0}$ component of the triplet scalar field $\Phi$ from the lagrangian above up to
$\frac{v^{4}}{f^{2}}$ order.
Those operators in Eq. (\ref{eff}) induce corrections to the masses of $W$ and $Z$ bosons after the scalar fields get VEVs.
After  $h$ and $\varphi$ get VEVs
 \begin{equation}
 <h>=\frac{v}{\sqrt{2}},
 \end{equation}
 \begin{equation}
 <\varphi>={v'},
 \end{equation}
 we obtain the masses of $W$ and $Z$ bosons and fermi constant $G_{F}$,
 which are presented in terms of the model
 parameters as follows;
 \begin{eqnarray}
 M_{_W}^{2}&=&g^{2}\frac{v^{2}}{4}\left(1+\frac{(s^{4}+6s^{2}c^{2}+c^{4})v^{2}}{4f^{2}}
 +4\frac{v^{'2}}{v^{2}}\right)~, \label{mw}\\
 M_{_Z}^{2}&=&(g^{2}+g^{'2})\frac{v^{2}}{4}\left(1+\frac{(s^{4}+6s^{2}c^{2}+c^{4})v^{2}}{4f^{2}}
                -\frac{5(s^{'2}-c^{'2})^{2}v^{2}}{4f^{2}}+8\frac{v^{'2}}{v^{2}}\right)~, \label{mz}\\
 \frac{1}{G_{F}}&=&\sqrt{2} v^{2}\left(1+\frac{v^{2}}{4f^{2}}+4\frac{v^{'2}}{v^{2}} \right)~.
\end{eqnarray}

Now, let us relate the model parameters to observables by using the
precision experimental values of $\alpha(M^{2}_{_Z}), M_{_Z}$ and
$G_F$ as inputs. {}From the standard definition of the weak mixing
angle $\sin \theta_0$ around the $Z$ pole given as
follows~\cite{Peskin:1991sw} ,
\begin{eqnarray}
\sin^{2}\theta_{0} \cos^{2}\theta_{0}&=&\frac{\pi
\alpha(M^{2}_{_Z})}{\sqrt{2}G_{F}M^{2}_{_Z}}~, \\
\sin^{2}\theta_{0}&=&0.23108\pm 0.00005~,
\end{eqnarray}
where $\alpha(M^{2}_{_Z})^{-1}=128.91\pm 0.02$  is the running SM
fine-structure constant evaluated at $M_{_Z}$~\cite{PDG}, we see
that the mixing angle $\sin\theta_W$ is related to  $\sin \theta_0$
through the relation,
\begin{eqnarray}
s^{2}_{0}&=&s^{2}_{W}+\delta
s^{2}_{W}=s^{2}_{W}-\frac{s^{2}_{W}c^{2}_{W}}{c^{2}_{W}-s^{2}_{W}}\left[\frac{\delta
G_{F}}{G_{F}}+\frac{\delta{M^{2}_{_Z}}}{M^{2}_{_Z}}-\frac{\delta \alpha}{\alpha}\right]~\nonumber \\
&=&
s^{2}_{W}-\frac{s^{2}_{W}c^{2}_{W}}{c^{2}_{W}-{s^{2}_{W}}}
\left[~4\Delta^{'}
+\Delta\left(-\frac{5}{4}+c^{2}(1-c^{2})+5c^{'}(1-c^{'2})\right)~\right]~,
\label{ds0}
\end{eqnarray}
where
\begin{equation}
\Delta = \frac{v^{2}}{f^{2}}~,~~~~~~
\Delta^{'}=\frac{v^{'2}}{v^{2}}~.
\end{equation}
Here, we omitted the $\delta \alpha$ term since there is no $\alpha$ correction.
Using the relations Eqs. (\ref{mw},\ref{mz},\ref{ds0}), we obtain
\begin{eqnarray}
\frac{M_{_W}^{2}}{M^{2}_{_Z}}-c_{0}^{2}=\frac{c^{2}_{W}}{c^{2}_{W}-s^{2}_{W}}
\left[~\Delta
\left(\frac{5}{4}c^{2}_{W}-s^{2}_{W}(c^{2}-c^{4})-5c^{2}_{W}(c^{'2}-c^{'4})\right)
-\Delta^{'}4c^{2}_{W}\right]~.
\end{eqnarray}
Finally we can get the form of $M_{_W}$ as a function of $c, c^{'}, f$, after substituting
the numerical value of $s_{0}$, as
\begin{eqnarray}
M_{_W}(c,c^{'},f) = (M_{_W})_{\rm SM}~
[1+\Delta \cdot {\cal G}(c,c^{'},f) + \Delta^{'} \cdot {\cal H}(c,c^{'},f)]~,
\end{eqnarray}
and for  $f \geq 4$ TeV, approximately
\begin{eqnarray}
M_{_W}  \simeq (M_{_W})_{\rm SM}~
[1+\Delta(0.89-0.21c^{2}+0.21c^{4}-3.6c^{'2}+3.6c^{'4})-2.9\Delta^{'}]~.
\label{wmass}
\end{eqnarray}
Therefore, it is reasonable that the $W$ boson mass $M_{_W}$ is decomposed into
the SM contribution $(M_{_W})_{\rm SM}$ and the shift due to new
tree-level contributions in the LHM~.
 \begin{figure}
\centering
\includegraphics[width=0.60\textwidth]{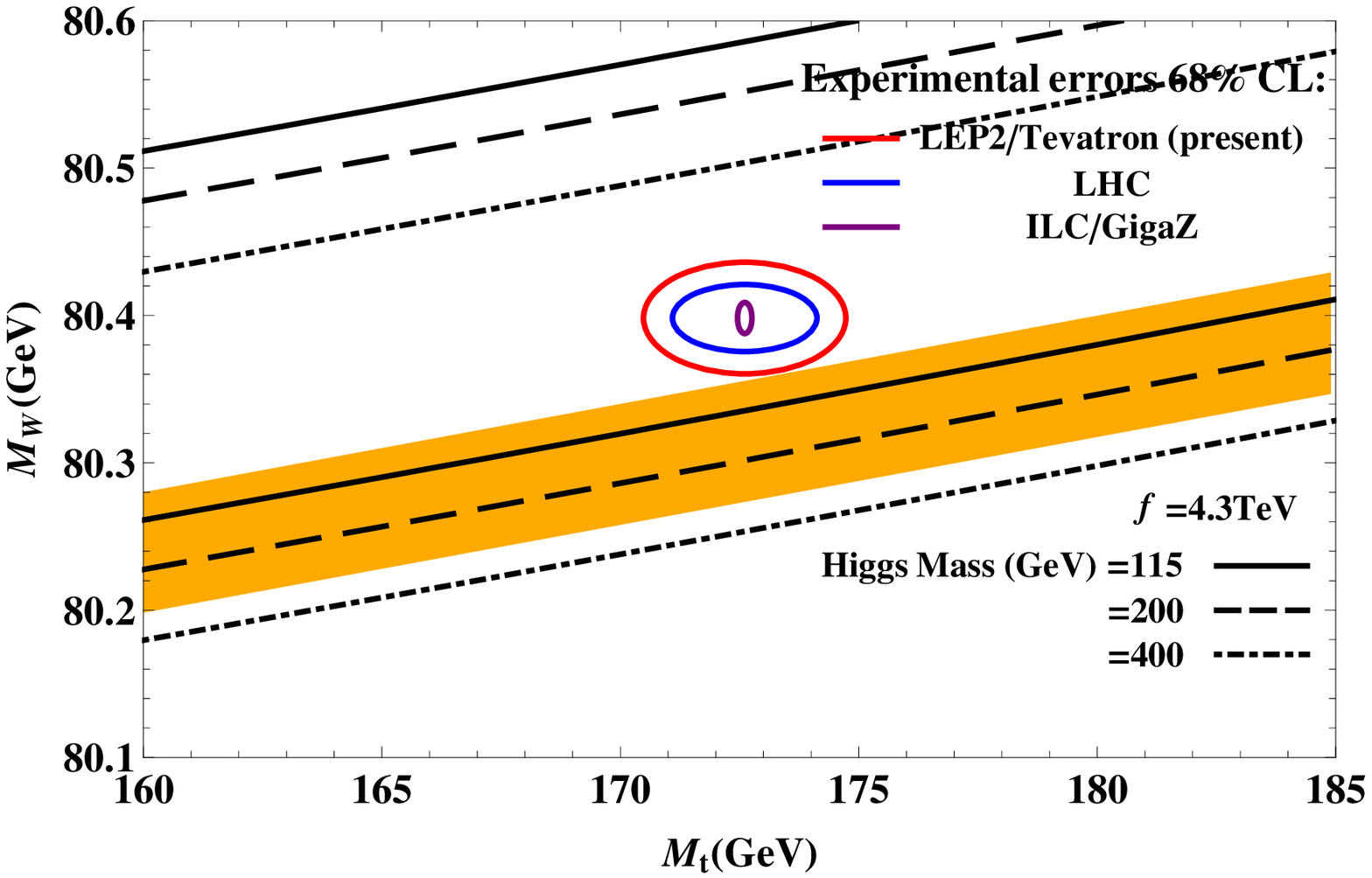}
\includegraphics[width=0.60\textwidth]{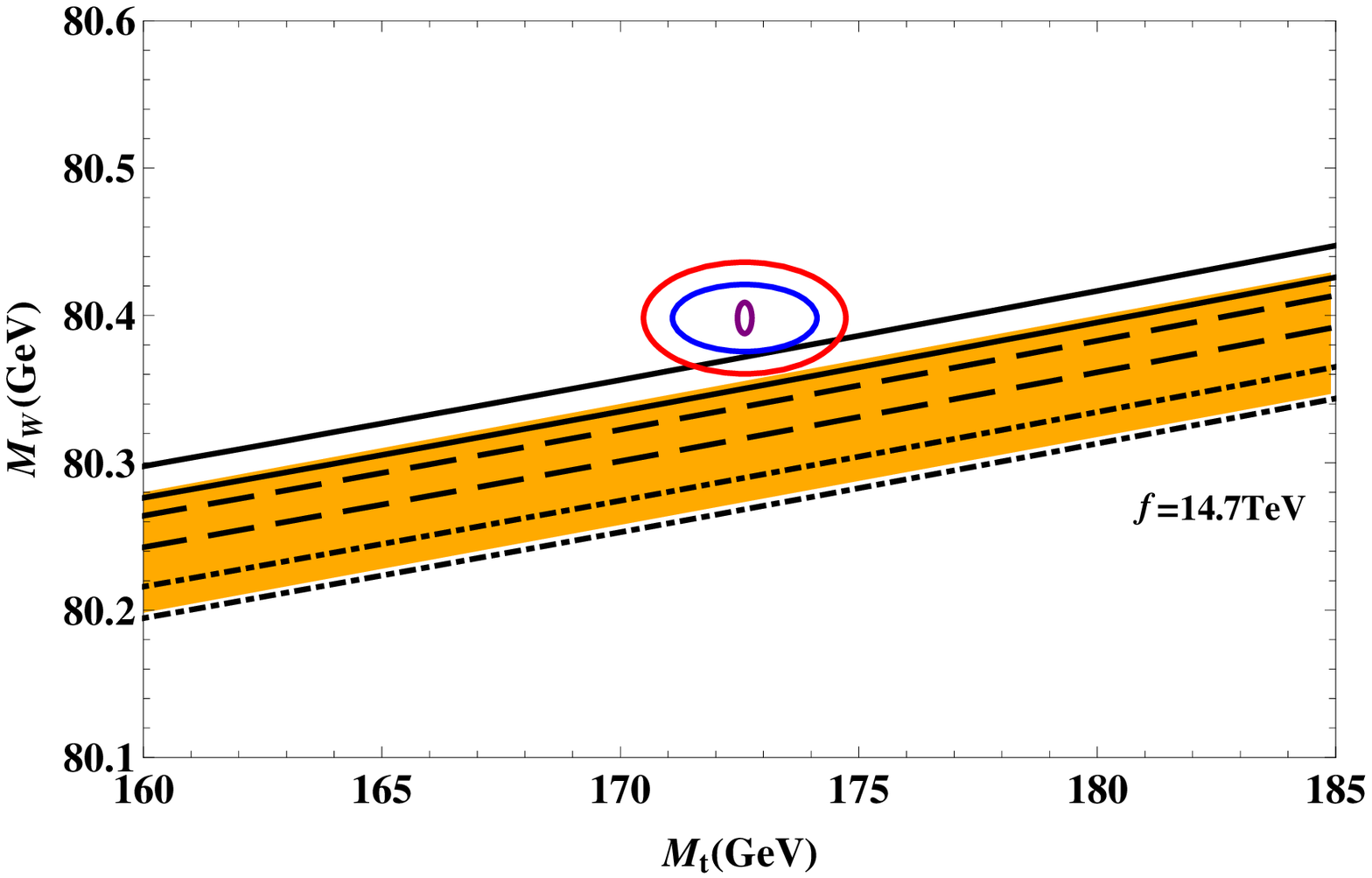}
\includegraphics[width=0.60\textwidth]{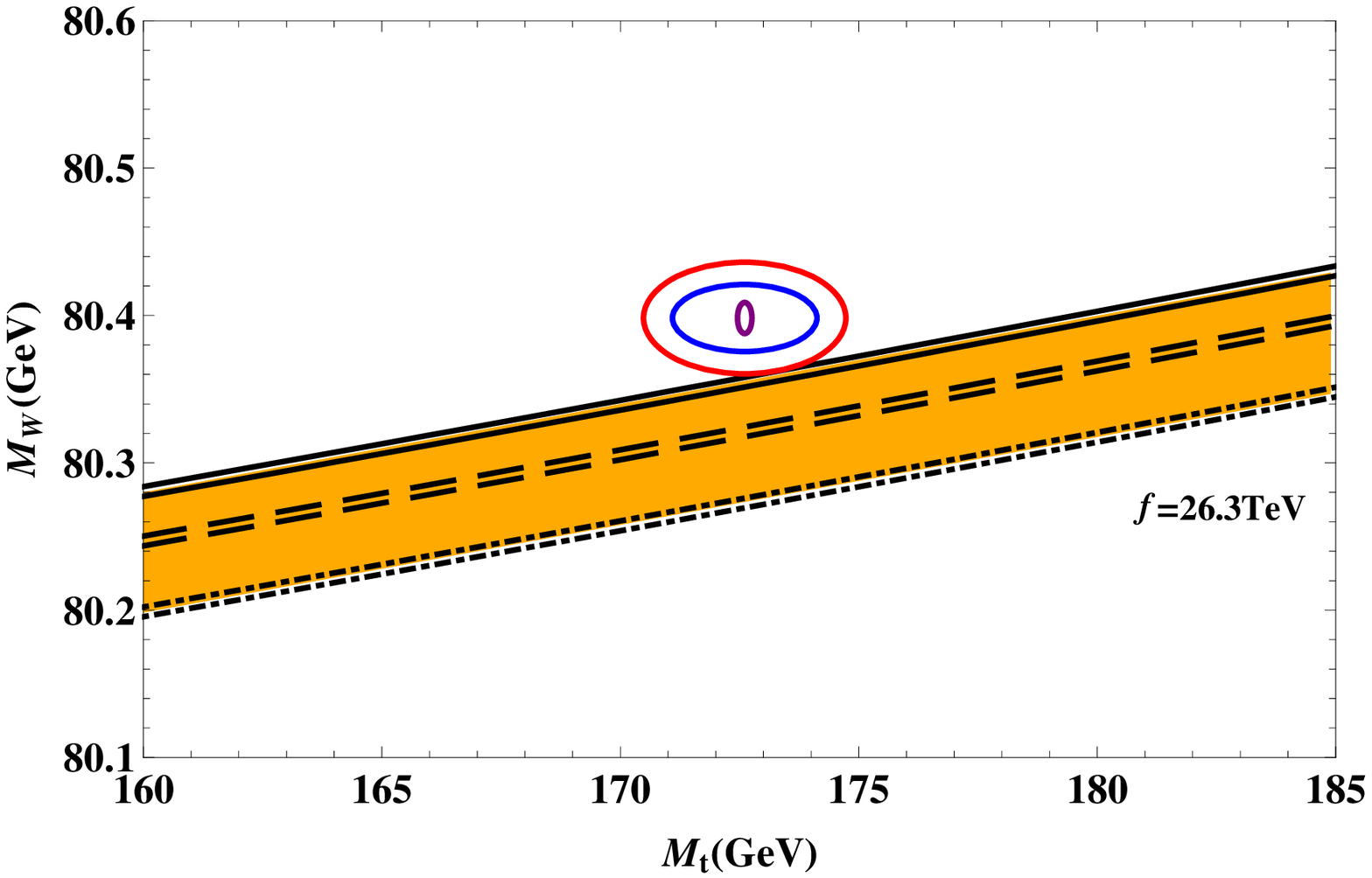}
\caption{Plots represent maximum (upper line) and minimum (lower
line) values of $M_{_W}$ as a function of $M_{t}$ in the SM (orange
colored band) and the LHM with $f=4.3,~ 14.7$ and $26.3$ TeV, where
solid, dashed and dot-dashed lines correspond to $M_{h}=115,~ 200$
and $400$ GeV, respectively. Red, blue and purple ellipses
correspond to the current measurements
~\cite{Alcaraz:2007ri,Group:2008nq}, prospective measurements at the
LHC~\cite{Beneke:2000hk,Haywood:1999qg},
 and at the ILC with GigaZ~\cite{Baur:2001yp,Abe:2001nnb} at the 68 $\%$ confidence level, respectively.}
\end{figure}

To compare the prediction of $W$ boson mass in the LHM with the
current measurements of $M_{_W}$ and $M_{t}$, we first compute the
SM contribution of the $W$-boson mass, $(M_{_W})_{\rm SM}$ by using
the fortran program package $\rm ZFITTER$ ~\cite{Arbuzov:2005ma}, in
which two and three loop corrections are included. In the numerical
estimation of $(M_{_W})_{\rm SM}$,
we take the five parameters, hadronic correction to the QED coupling
$\Delta^{(5)}_{h}$, the QCD coupling $\alpha_{\rm s}$, the $Z$ boson
mass $M_{_Z}$,  the top quark mass $M_{t}$ and  the Higgs mass
$M_{h}$, as input parameters. For their numerical values, we take
$\Delta^{(5)}_{h}=0.02802(15)$, $\alpha_{\rm s}=0.1216(17)$,
$M_{_Z}=91.1874(21)$ GeV. For the input values of $M_{t}$ and
$M_{h}$, we consider the ranges $160 \leq M_{t} \leq 185$ GeV and
$115 \leq M_{h} \leq 400$ GeV, respectively, in order to see how the
prediction of $M_{_W}$ is correlated with $M_{t}$ and $M_{h}$. Here,
the lower limit of $M_{h}$ is adopted from the direct search at LEP
\cite{Barate:2003sz}. As one can see from Eq. (\ref{wmass}), the
part of the shift of $M_{_W}$ from $(M_{_W})_{\rm SM}$ due to new
contributions of the LHM depends on the parameters $c$,
$c^{\prime}$, $\Delta$, and $\Delta^{\prime}$. For the sake of
simplicity, we set the triplet VEV $v'$ to be zero. We note in fact
that this triplet VEV  turns out to generate sub-leading
contributions~\cite{Csaki:2002qg}. Thus, in this work, the model
dependent input parameters are $c$, $c^{\prime}$ and $\Delta$. Among
them, the parameters $c$ and $c^{\prime}$ are restricted to be
$-1\leq c(c^{\prime}) \leq 1$ and $\Delta$ should be much less than
one. For example, if we take $f\simeq 1$ TeV, then $\Delta \simeq 0.06$.

Based on the formulae for $M_{_W}$ given in Eq. (\ref{wmass}) and
taking appropriate numerical values for the input parameters
including $M_{t}$ and $M_{h}$, we finally obtain the prediction of
$M_{_W}$ in the LHM. It is worthwhile to notice that there exist
upper and lower limits for the prediction of $M_{_W}$ for a fixed
parameter set ($M_{t}, M_{h}$, $\Delta$) due to the restriction of
the mixing parameters $c$ and $c^{\prime}$. As one can expect, the
gap between the upper and lower limits for the prediction of
$M_{_W}$ for a given $M_{h}$ gets smaller as the value of $f$
increases.

In Fig. 1, we show the predictions of $M_{_W}$ in the SM and the LHM
with $f$=4.3, 14.7 and 26.3 TeV as a function of $M_{t}$. The reason
why we take those particular values of $f$ will become clear from
the discussions presented below. The orange colored bands in Fig. 1 indicate
the SM prediction of $M_{_W}$ for $115~{\rm GeV} \leq M_{h} \leq
400~{\rm GeV}~$. As is well known, the SM prediction of $M_{_W}$ for
a fixed $M_{t}$ gets smaller as $M_{h}$ increases, so the upper and
lower limits for the orange bands correspond to $M_{h}=115$ GeV and
$M_{h}=400$ GeV, respectively. Similarly, the solid, dashed and
dotted lines correspond to the upper and lower limits for the
prediction of $M_{_W}$ for $M_{h}$=115, 200 and 400 GeV,
respectively in the LHM. In the center of each panel, the red
ellipse represents the current experimental
results of LEP2/Tevatron, $M_{_W}=80.398\pm 0.025$~\cite{Alcaraz:2007ri} and
$M_{t}=172.6\pm 1.4$ GeV~\cite{Group:2008nq},
the blue and purple represent the same central values with prospective uncertainties for
$M_{_W}$ and $M_{t}$ as the current ones achievable at the LHC
~\cite{Beneke:2000hk,Haywood:1999qg},
\begin{equation}
\delta M_{_W}=15 ~{\rm{MeV}}~,~~~~~~\delta M_{t}=1.0 ~\rm{GeV}~,
\end{equation}
 and at the ILC/GigaZ~\cite{Baur:2001yp,Abe:2001nnb},
\begin{equation}
\delta M_{_W}=7 ~{\rm{MeV}}~~ ,~~~~~~\delta M_{t}=0.1 ~\rm{GeV}~,
\end{equation} at 1$\sigma$ CL, respectively.
 It is likely that the
current experimental data for $M_{_W}$ and $M_{t}$ disfavors the SM
prediction of $M_{_W}$ at 1 $\sigma $ CL.  As shown in Fig. 1, if
the future measurements of $M_{_W}$ and $M_{t}$ at the LHC and ILC
would be done like the blue and purple ellipses, it could serve as a
hint for the existence of new physics beyond the SM.

We see from Fig. 1 that in the case of $f=4.3$ TeV, the predictions
of $M_{_W}$ in the LHM for the given range of $M_{h}$ cover the
whole regions of the ellipses. However, in the case of $f=14.7$ TeV,
the $1\sigma$ ellipse for the current measurements of $M_{_W}$ and
$M_{t}$ is consistent with the prediction of $M_{_W}$ for
$M_{h}=115$ GeV but appears to be inconsistent with the predictions
for larger values of $M_{h}$. In our numerical estimation, we have
observed that the predictions of $M_{_W}$ for $f \gtrsim 14.7$ TeV
deviate from the $1\sigma$ ellipse for the prospective measurements
of $M_{_W}$ and $M_{t}$  achievable at the LHC, and thus $f=14.7$
TeV could be regarded as an upper bound on $f$ in the LHM in the LHC
era. In the case of $f=26.3$ TeV, even the $1\sigma$ ellipse for the
current measurements of $M_{_W}$ and $M_{t}$ starts to deviate from
the whole region of the prediction for $M_{_W}$ in the LHM,  and
it is almost the same as the SM prediction of $M_W$. Thus, $f=26.3$
TeV can be regarded as the current upper bound on the symmetry
breaking scale  in the LHM.

\begin{figure}[t]
\centering
\includegraphics[width=0.7\textwidth]{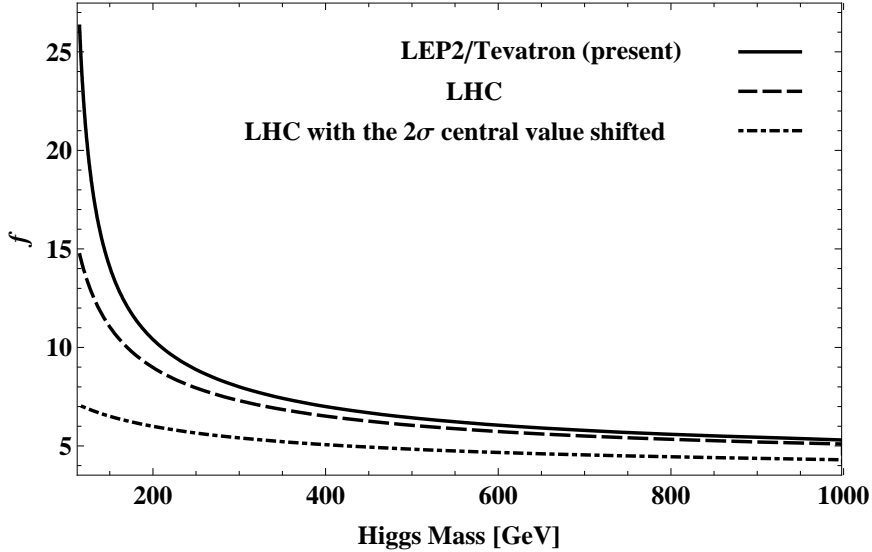}
\caption{Plots represent the upper bound on $f$ as a function of the
Higgs mass $M_{h}$. The soild, dashed and dot-dashed curves
correspond to the cases of the ellipses obtained from the current
data, the LHC prospect with the same central values as the current
ones, and the LHC prospect with different central values
($i.e.$ $M_{t}=169.8$ GeV, $M_{_W}=80.448$ deviated from the current values
by $2\sigma$), respectively. }
\end{figure}

It is worthwhile to notice that the upper bound on $f$ obtained
above is closely related with the current lower limit on the Higgs
mass $M_{h}=114$ GeV. If the Higgs boson with $M_{h}>114$ GeV
is discovered  or the lower limit on $M_{h}$  is increased  in
the future, the upper bound on $f$ will be decreased to the values lower than
$f=14.7$ TeV.
%
%

In Fig.  2, we show how the upper bound on $f$ depends on the Higgs
mass. The solid, dashed and dot-dashed curves correspond to the
cases of the ellipses obtained from the current data, the LHC
prospect with the same central values as the current ones, and the
LHC prospect with different central values ($i.e.$ $M_{t}=169.8$ GeV,
$M_{_W}=80.448$ GeV corresponding to $2\sigma$ deviation from the
present central values), respectively. In this plot we see that as
$M_{h}$ decreases, the upper bound on $f$ rapidly increases. If a
light Higgs boson with mass, for example, roughly $M_{h}\sim$ 200
GeV is observed at the LHC, the results in Fig. 2 indicate
that the value of $f$ will be below about 9 TeV. On the other
hand, if the Higgs mass is measured to be rather heavy
($M_{h}$ $\sim$ 800 GeV), $f$ will be below 5.3 TeV. Here, note
that we allow $M_{h}$ to be up to 1 TeV because of the unitarity of the
longitudinal $W_{\rm L}-W_{\rm L}$ scattering
amplitude~\cite{Lee:1977eg}. Thus, taking the Higgs mass
$M_{h}=1$  TeV, the upper bound on $f$ lowers down to 5.0 (4.3) TeV
for solid (dashed) curve. Therefore, as the upper bound on $f$ gets
increased, the allowed Higgs mass in the context of the LHM gets
smaller. It is also worthwhile to see that the  shift of the central
values for $M_{_W}$ and $M_{t}$ while keeping the same uncertainties,
the case corresponding to the dot-dashed curve, lowers the upper
bound on $f$. In addition, as expected, the reduction of the
uncertainties in future experiments such as the LHC and ILC must
lower the upper bound on $f$, too. It is interesting to notice that
there exists a lower bound on $f$, $f\geq 4$ TeV at $95\%$ CL,
coming from the global fit to electroweak precision data
\cite{Csaki:2002qg}, and for certain variations of the LHM there
exists a parameter space which can bring the $f$ value as low as $1 \sim
2$ TeV by changing the U(1)$\times$U(1) charge assignments of the SM
fermions~\cite{Csaki:2003si}.
 Combining the lower bound on $f$ from the global
fit together with the upper bound estimated  here,
 we can narrow down the range of the symmetry breaking scale $f$.
 Such a narrow range of $f$ may be useful to investigate the effects of the LHM, which can be probed at the LHC.


\section{Constraints on the mixing parameters and heavy gauge boson masses }

Let us investigate how the allowed regions of the mixing parameters
$c$ and $c'$ in the LHM can be extracted from comparison with
experimental results. Bearing in mind that both mixing parameters
$c$ and $c'$ have finite domain ( -1 $\leq c$, $c'$ $\leq$ 1 ), we
first scan all possible points of $c$ and $c'$ on calculating
$M_{_W}$. We then pick up the values of $c$ and $c'$ for which the
prediction of $M_{_W}$ for fixed values of $M_{h}$ and $f$ is
consistent with the 1$\sigma$ ellipse for the current measurements
of $M_{_W}$ and $M_{t}$. In this way, we obtain the allowed regions
of the mixing parameters $c$ and $c^{\prime}$. For our numerical
calculation, we take several cases, $f$=1, 4, 5 and 7.

Fig.  3  presents the allowed regions for $c$ and $c'$ for given
values of $f$. In each panel, the colored bands correspond to the
allowed regions of the parameter space ($c$ and $c^{'}$) for
$M_{h}$= 115, 200, 300 and 400 GeV, respectively.
 It is interesting to see that
the mixing parameter $c'$ is rather strongly constrained whereas $c$
is not constrained at all. This is because the prediction of
$M_{_W}$ is much more sensitive to $c^{'}$ rather than $c$ for a
given parameter set as can be seen from Eq. (\ref{wmass}). In the
case of $f$=1 TeV, the gap of each band is very narrow
compared with those for other cases. And for the case with $f$ smaller than 1 TeV
this feature  almost does not change at all.
 There also exist common forbidden parameter regions around $c^{'}\sim 0.7$ for all values of $f$. The
forbidden region is expanded as $f$ increases. In fact, the size of
$\Delta$ gets larger as $f$ decreases, so for the realm of small $f$,
small change of $c'$ leads to rather large change of $M_{_W}$,
whereas the sensitivity of $M_{h}$ and $M_{t}$ through
$(M_{_W})_{\rm SM}$ to $M_{_W}$ is not substantial. For $f\gtrsim 4$
TeV, the allowed regions of $c'$ appears to be expanded as $f$
increases, and they include very small $c^{'}$ for large values of
$M_{h}$. This is because the value of $\Delta$ gets smaller as $f$
increases, so the sensitivity of $c'$ to $M_{_W}$ becomes weaker
whereas that of $M_{h}$ to $M_{_W}$ becomes stronger. For a fixed
value of $M_{h}$, the boundaries of the allowed region for $c'$ are
extended as $f$ increases. For the case of $f$=7 TeV,  as can be
seen from Fig. 3, there is no allowed region of $c$ and $c'$ for
$M_{h}\gtrsim 400$ GeV.  This can be regarded as an upper limit
of $M_{h}$ along with $f$ allowed in the context of the LHM.

\begin{figure}[t]
\includegraphics[width=1.0\textwidth]{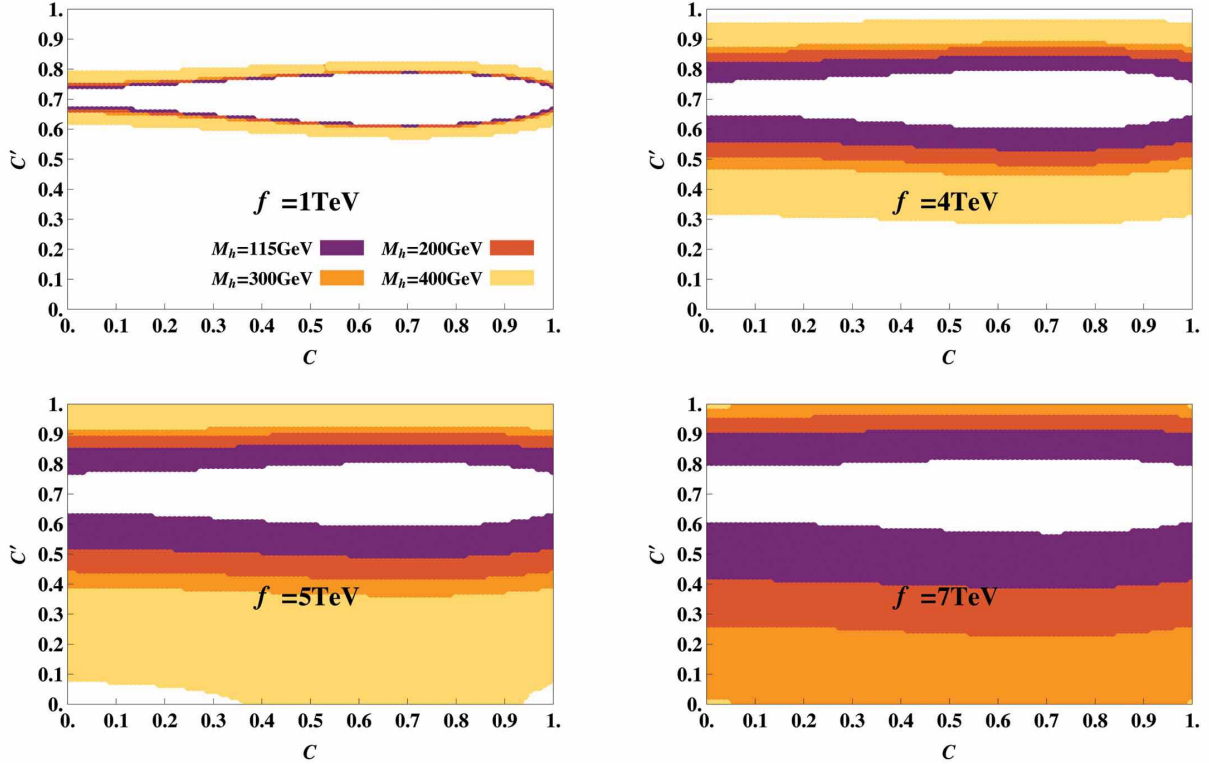}
\caption{We plot the allowed region of the parameter space ($c$ and
$c'$) for $f$=1, 4, 5 and 7 TeV, where the four colors correspond
to $M_{h}$=115, 200, 300 and 400 GeV, respectively.}
\end{figure}

\begin{table}[h]
\begin{center}
\caption{  The allowed regions of the mixing parameter $c'$ are
presented for $f=1,2$ and $4$ TeV and $M_h=115, 200, 300$ and $400$,
respectively. Note that there are two allowed regions for each $f$
and $M_{h}$. Note also that there are  common forbidden regions
($0.69< c' <0.73$).\\}
\begin{tabular}{ccccc}
\hline
\multicolumn{5}{|c|}{Constraints on the mixing parameter $c^{'}$ and
the mass of the heavy gauge boson $B^{'}$} \\
\hline
~ $f$ value ~&~ $M_{h}$ ~&~ Bottom region ~&~ Top region ~&~ expected $B^{'}$ mass\\
\hline
~~1 TeV~    &~ 115 GeV~&~ $0.60 \sim 0.69$~ & $0.73 \sim 0.80$~~& $159.4 \sim 165.9$,~ $159.6 \sim 165.9$ GeV\\ 
            &~ 200 GeV~&~ $0.60 \sim 0.67$~ & $0.75 \sim 0.80$~~& $160.1 \sim 165.9$,~ $160.5 \sim 165.9$ GeV\\ 
            &~ 300 GeV~&~ $0.59 \sim 0.66$~ & $0.75 \sim 0.80$~~& $160.6 \sim 167.1$,~ $160.5 \sim 165.9$ GeV\\ 
            &~ 400 GeV~&~ $0.59 \sim 0.65$~ & $0.76 \sim 0.81$~~& $161.2 \sim 167.1$,~ $161.2 \sim 167.6$ GeV\\ 
\hline
~~2 TeV~    &~ 115 GeV~&~ $0.56 \sim 0.67$~ & $0.74 \sim 0.82$~~& $320.1 \sim 343.2$,~ $319.9 \sim 339.3$ GeV\\
            &~ 200 GeV~&~ $0.55 \sim 0.63$~ & $0.78 \sim 0.84$~~& $325.5 \sim 346.7$,~ $326.2 \sim 349.4$ GeV\\ 
            &~ 300 GeV~&~ $0.53 \sim 0.61$~ & $0.80 \sim 0.84$~~& $329.4 \sim 354.3$,~ $331.7 \sim 349.4$ GeV\\ 
            &~ 400 GeV~&~ $0.52 \sim 0.59$~ & $0.81 \sim 0.85$~~& $334.3 \sim 358.5$,~ $335.2 \sim 355.6$ GeV\\ 
\hline
~~4 TeV~    &~ 115 GeV~&~ $0.42 \sim 0.60$~ & $0.80 \sim 0.90$~~& $663.4 \sim  835.5$,~ $637.9 \sim  787.5$ GeV\\
            &~ 200 GeV~&~ $0.36 \sim 0.53$~ & $0.85 \sim 0.93$~~& $708.6 \sim  948.2$,~ $690.0 \sim  903.8$ GeV\\ 
            &~ 300 GeV~&~ $0.32 \sim 0.48$~ & $0.88 \sim 0.95$~~& $756.3 \sim 1050.4$,~ $739.2 \sim 963.4$ GeV\\ 
            &~ 400 GeV~&~ $0.28 \sim 0.44$~ & $0.90 \sim 0.96$~~& $806.0 \sim 1148.7$,~ $811.7 \sim 1148.7$ GeV\\ 
\end{tabular}
\end{center}
\end{table}

The constraint on $c^{\prime}$ obtained above enables us to estimate
the masses of  heavy gauge bosons in the LHM. The masses of the
heavy gauge bosons $W^{a'}_{\mu}$ and $B^{'}_{\mu}$ are given in
terms of mixing parameters by
\begin{equation}
M_{W^{'}}=\frac{g}{2sc}f~\geq
gf~,~~M_{B^{'}}=\frac{g^{'}}{2\sqrt{5}s^{'}c^{'}}f~\geq
\frac{g^{'}f}{\sqrt{5}}~.
\end{equation}
In addition to those derived lower bounds on the masses of heavy gauge
bosons, we can constrain the size of $M_{B^{'}}$ further by imposing
the constraint on $c^{\prime}$ obtained above.

In Table I, we present the predictions of  $M_{B^{'}}$ for several
combinations of $f$ and $M_{h}$ along with the constraints on
$c^{\prime}$. As the value of $f$ decreases, $M_{B^{'}}$ is
predicted to get smaller and the theoretical uncertainty
gets narrower.
In the light of search for new physics, that is a very important
implication for the verification of the validity of the LHM when we
get to probe or even observe a certain signal for new
additional gauge bosons at future colliders.
\\


In conclusion, based on the prediction of $M_{_W}$ in the LHM, we
have compared it with the current and prospective measurements of
$M_{_W}$ and $M_{t}$,
 and found that the current values and accuracy of $M_{_W}$ and $M_{t}$
measurements tend to favor the LHM over the SM, although the most
recent electroweak data may appear to be consistent with the SM
prediction.
We have found that the predictions of $M_{_W}$ in the LHM for
$f\gtrsim 26.3$ TeV deviate from the realm of the 1$\sigma$ ellipse
for the measurements of $M_{_W}$ and $M_{t}$, and thus $f=26.3$ TeV
can be regarded as the upper bound on $f$. We have discussed how the
upper bound on $f$ depends on the Higgs boson mass. As $M_{h}$
decreases, the upper bound on $f$ rapidly increases.
We have examined how the parameters $c$ and $c'$ can be constrained
by comparing the prediction of $M_{_W}$ with the current precision
measurements of $M_{_W}$ and $M_{t}$. For a given parameter set, it
turns out that $c'$ is strongly constrained for small $f$
whereas $c$ is not constrained at all. We have studied how the mass
of the heavy gauge boson $M_{B^{'}}$ in the LHM can be extracted
from the constraint on $c^{\prime}$ for a given value of $f$. We
anticipate that more precision data for $M_{_W}$ and $M_{t}$ as well
as even discovery of the Higgs boson at the LHC would give the LHM
even more preference and provide a decisive clue on the evidence of the LHM.
\\

\centerline{\bf ACKNOWLEDGEMENTS} \noindent We thank
G. Cvetic for careful reading of the manuscript and his
valuable comments.
JBP was supported  by
the Korea Research Foundation Grant funded by the Korean Government
(MOEHRD) No. KRF-2005-070-C00030. CSK was supported in part by
CHEP-SRC Program and in part by the Korea Research Foundation Grant
funded by the Korean Government (MOEHRD) No. KRF-2005-070-C00030.
SKK was supported by KRF Grant funded by the Korean Government
(MOEHRD) No. KRF-2006-003-C00069.


\end{document}